\documentclass[conference,10pt]{IEEEtran}
\usepackage{epsfig,rotating,setspace,latexsym,amsmath,epsf,amssymb,amsfonts,bm,subfigure,epstopdf,cite,authblk,bbm,mathtools,amsthm,algorithm,color,systeme,mathtools,amsmath,derivative,bigints}

\usepackage[utf8]{inputenc}
\usepackage[T1]{fontenc}
\usepackage[thinc]{esdiff}
\usepackage[noend]{algpseudocode}
\theoremstyle{plain}

\newtheorem{theorem}{Theorem}
\newtheorem{lemma}{Lemma}

\newtheorem{definition}{Definition}

\algrenewcommand\algorithmicforall{\textbf{foreach}}
\algrenewcommand\algorithmicindent{.8em}
\usepackage{tabularx}

\DeclareMathOperator*{\avg}{avg}

\IEEEoverridecommandlockouts
\allowdisplaybreaks

\begin{document}


\title{Preemptive Scheduling for Age of Job Minimization in Task-Specific Machine Networks}

\author{Subhankar Banerjee \qquad Sennur Ulukus\\
\normalsize Department of Electrical and Computer Engineering\\
\normalsize University of Maryland, College Park, MD 20742\\
\normalsize \emph{sbanerje@umd.edu} \qquad \emph{ulukus@umd.edu} }

\maketitle

\begin{abstract}
    We consider a time-slotted job-assignment system consisting of a central server, $N$ task-specific networks of machines, and multiple users. Each network specializes in executing a distinct type of task. Users stochastically generate jobs of various types and forward them to the central server, which routes each job to the appropriate network of machines. Due to resource constraints, the server cannot serve all users' jobs simultaneously, which motivates the design of scheduling policies with possible preemption. To evaluate scheduling performance, we introduce a novel timeliness metric, the \emph{age of job}, inspired by the well-known metric, the age of information. We study the problem of minimizing the long-term weighted average age of job. We first propose a max-weight policy by minimizing the one-step Lyapunov drift and then derive the Whittle index (WI) policy when the job completion times of the networks of machines follow geometric distributions. For general job completion time distributions, we introduce a Whittle index with max-weight fallback (WIMWF) policy. We also investigate the Net-gain maximization (NGM) policy. Numerically, we show that the proposed WIMWF policy achieves the best performance in the general job completion time setting. We also observe a scaling trend: two different max-weight policies can outperform the NGM policy in small systems, whereas the NGM policy improves as we scale the system size and becomes asymptotically better than max-weight policies. For geometric service times, the WI policy yields the lowest age across all considered system sizes.
\end{abstract}

\section{Introduction}\label{sec:intro}
In applications such as intelligence, surveillance, and other mission-critical control tasks, job offloading to edge computing devices has become a key mechanism for handling latency-sensitive workloads~\cite{liu2019dynamic}. Different applications and users require different types of computations and have heterogeneous latency requirements. This naturally leads to a system architecture with multiple task-specific networks of edge computing devices, where each network is specialized for a particular class of jobs. To ensure that each task is routed to the appropriate network, we assume the presence of a central server that collects task requests from all users or applications and forwards them to the corresponding edge network.

This architecture is closely aligned with the recent developments in task-specific large language models (LLMs). Training a single large, general-purpose LLM typically requires massive datasets, long training times, and substantial compute resources; while such models perform well across diverse tasks, the associated training and deployment costs can become a bottleneck. In contrast, smaller-parameter and task-specialized LLMs can be trained more efficiently on narrower domains, requiring less data and time, and often outperforming a large general-purpose LLM on the specific tasks for which they are designed; see, for example,~\cite{yangspecialized}. The model considered in our work naturally captures the problem of routing tasks across networks of task-specific LLMs, while remaining general enough to encompass other task-specific edge computing services and architectures.

\begin{figure}[t]
    \centerline{\includegraphics[width = 0.8\columnwidth]{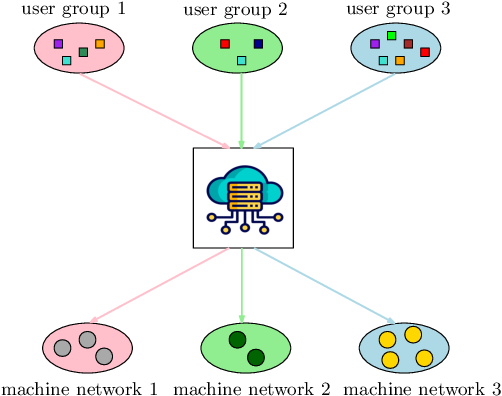}}
    \caption{Colored squares represent users. The three ellipses at the top represent user groups, each associated with a specific task. A user may belong to multiple groups; for example, the red user belongs to both the green and blue groups. Users submit their requests to the central server, which routes them to the corresponding network of task-specific machines, shown in the bottom ellipses, where machines are represented by circles. For instance, the gray machines serve the task associated with the pink group. [Figure credit: central processor icon made by RaftelDesign at www.flaticon.com.]}
    \label{fig:1}
\end{figure}

In this work, we assume that there are $N$ groups of users and $M$ users in total; the $i$th group consists of $M_{i}$ users, and every user in a given group is interested in the same specific application. At each slot, user~$j$ from group~$i$ submits a job request to the central server with probability $p_{j}^{i}>0$. For the $i$th group of users, there is a corresponding $i$th network of edge computing devices specialized to perform the task that group~$i$ is interested in. We impose resource constraints: the $i$th edge network can execute at most $\bar{M}_{i}$ jobs concurrently, and the central server can process at most $\bar{M}$ jobs concurrently. Thus, the server must employ a scheduling algorithm that decides which jobs to serve in every time slot. At a given slot, if the server decides to execute a job for user $j$ in group~$i$, it routes that job to edge network~$i$, which then executes the job according to its service time distribution. In this paper, we use the terms \emph{edge computing device} and \emph{machine} interchangeably, and likewise \emph{edge network} and \emph{machine network}. We pictorially represent our system in Fig.~\ref{fig:1}.

In this work, we allow \emph{preemption} of in-service jobs in the sense of \emph{preemptive-resume} service: an ongoing job can be interrupted to serve a more latency-sensitive job, and the progress of the interrupted job is stored in a buffer and later resumed from where it left off, rather than being discarded. When a user submits a job to the central server, our goal is to serve that job in a timely manner. Motivated by the well-known metric \emph{age of information}, we introduce a novel timeliness metric, the \emph{age of job}. To capture the relative importance of timeliness and latency across jobs, we weight the age of job for each user in every group. Our aim is to design near-optimal scheduling policies for the server that minimize the long-term average weighted age of job in the system.

A natural question is why we consider an age-type penalty instead of simply minimizing the delay between a user submitting a job and an edge server executing it. In a subsequent work, we show that if we optimize only delay instead of our age of job penalty, then, under an optimal scheduling policy, the server may never schedule certain users, which is undesirable. Also, our proposed metric age of job is similar to the \emph{age of job completion} introduced in~\cite{mitrolaris2025age}, except for a small difference that we describe in Section~\ref{sec:related-work}. Thus, similar to~\cite{mitrolaris2025age}, we argue that, minimizing the age of job corresponding to a user is equivalent to maximizing the number of jobs completed per unit time for that user.

We propose multiple scheduling policies. First, we derive a max-weight policy by minimizing the one-slot Lyapunov drift. We then formulate the scheduling problem as a restless multi-armed bandit (RMAB) problem. In the seminal paper~\cite{whittleindex1}, Whittle introduced an index-based policy to solve RMAB problems with binary actions, and since then many works have derived Whittle index policies under various settings; see, for example,~\cite{verlupewhittle}. A key challenge in applying the Whittle index is the need to prove that the underlying problem is indexable. For many systems, establishing indexability is difficult, and for this reason the \emph{Net-gain maximization} (NGM) policy, which does not require indexability, has also been studied~\cite{shisher2023learning, ornee2023context, chakraborty2024timely, chamoun2025edge}. For our model, when the service time of the edge computing devices follows a geometric distribution, we establish indexability and derive the corresponding Whittle index policy. For general service time distributions, however, we do not have the indexability property. In this case, inspired by the Whittle index framework, we propose a Whittle index with max-Weight fallback (WIMWF) policy. We also derive the NGM policy for our problem. We compare the performance of our proposed policies numerically in Section~\ref{sec:num_res}.

\section{Related Works}\label{sec:related-work}
Recently, job offloading to remote machines has been studied extensively; see, for example,~\cite{banerjee2025tracking, liyanaarachchi2025optimum, liyanaarachchi2025age, sariisik2025maximize,chamoun2025edge, chamoun2025mappo, mitrolaris2025age}. All of these works modeled each remote machine as a \emph{Markov machine}: when the machine was not serving a job assigned by the server, it could either be busy serving an internal job or idle, and its busy or idle state evolved as a Markov chain. As a result, a job submitted by the server could be dropped if the machine was busy with an internal job, and tracking the machine state became a central challenge. In~\cite{banerjee2025tracking}, the authors considered minimizing a tailored version of the \emph{age of incorrect information} together with a job-dropping penalty, while~\cite{chamoun2025edge} minimized job-dropping uncertainty in a network with multiple job dispatchers and multiple Markov machines. 

Because of the need to track the Markov machines, the works~\cite{mitrolaris2025age,chamoun2025mappo,liyanaarachchi2025optimum,sariisik2025maximize} assumed that the job-assigning entity (central server, job dispatcher, or resource allocator) periodically requested or sampled the state of each machine to obtain a better estimate of its state. In some of these works, a sampling cost was explicitly modeled: for example, in~\cite{chamoun2025mappo}, the authors optimized the differential gain between successful job completion and sampling cost, while in~\cite{mitrolaris2025age}, a novel \emph{age of job completion} metric was introduced and the average age of job completion was minimized together with a sampling cost under queue-stability conditions. Other works imposed a sampling constraint and studied a sampling-rate allocation problem~\cite{liyanaarachchi2025optimum,sariisik2025maximize}. In~\cite{liyanaarachchi2025optimum}, a resource allocator assigned jobs to a network of machines and tracked their states under a sampling constraint; to evaluate tracking performance, the authors used the well-known \emph{binary freshness} metric and introduced two additional metrics, \emph{false acceptance ratio} and \emph{false rejection ratio}. In~\cite{sariisik2025maximize}, the authors considered a network of exhausted workers whose efficiency depended on their internal states, and the resource allocator aimed to determine an optimal rate allocation scheme that maximized the probability that a worker operated in its most efficient state.

In contrast, in our model, a machine that is not serving a job assigned by the central server is simply idle, and there is no internal job process. Hence, once the server assigns a job to a machine, there is no uncertainty: the job will be executed and never dropped, and explicit tracking of machine states is not needed. We also optimize a different timeliness metric: we introduce the new metric \emph{age of job}. The closest metric to ours is the age of job completion proposed in~\cite{mitrolaris2025age}; however, unlike the age of job completion, the age of job increases only when a job for that user is present in the central server, and our networked architecture is more complex than the single Markov machine considered in~\cite{mitrolaris2025age}. Moreover, while~\cite{mitrolaris2025age} focused on a class of randomized policies, we consider several policy families, including the Whittle index policy, which is known to exhibit asymptotic optimality under certain technical conditions~\cite{verlupewhittle}. Finally, none of the works in~\cite{chamoun2025edge, banerjee2025tracking, liyanaarachchi2025optimum, liyanaarachchi2025age, sariisik2025maximize, chamoun2025mappo, mitrolaris2025age} allowed preemption of in-service jobs, whereas our model explicitly incorporates \emph{preemptive-resume} service, making our problem fundamentally different from these works.

In~\cite{turkmen2025balancing}, the authors considered a system with several task-specific networks of LLMs and multiple users that submitted binary queries to a central server. Although the system model was close to ours, they optimized the number of LLMs in each network to balance information accuracy and response delay, and preemption was not allowed. In~\cite{jeff2025optimal}, a two-mode job offloading system was considered, where jobs were either processed entirely in the cloud or partially at a local server and partially in the cloud, and the goal was to minimize the delay; here also, preemption was not considered.

In the age of information literature, preemptive transmission policies have been studied for various network models; see, for example,~\cite{najm2018status, yates2018age2, arafa2019aller, wang2019preempt, banerjee2024preempt, moltafet2025multi, dogan2021multi}. Some works derived closed-form age expressions under always-preemptive or probabilistically-preemptive policies~\cite{najm2018status, yates2018age2, dogan2021multi, moltafet2025multi}, while others posed optimization problems to identify optimal preemptive policies~\cite{arafa2019aller, wang2019preempt, banerjee2024preempt}. Our work is related in that we also design and analyze preemptive policies, but it differs in an important way. In most age-based preemption works, preemption typically implies discarding the in-service packet, whereas our model uses preemptive-resume service, in which an interrupted job is halted, its progress stored, and later resumed from where it left off rather than being discarded.

\section{System Model and Problem Formulation}
We denote the $i$th group of users by the set $\mathcal{M}_{i}$. We let $M_{i} = |\mathcal{M}_{i}|$. The user sets need not be disjoint, i.e., a single user may be associated with multiple tasks executed by different networks. Consequently, in general we have $\sum_{i=1}^{N} M_{i} \geq M$. At time $t$, we represent the job-generation event for user $j\in\mathcal{M}_{i}$ by an indicator random variable $a_{j}^{i}(t)$, where $a_{j}^{i}(t)=1$ implies that a job is generated, and $a_{j}^{i}(t)=0$ otherwise. For user $j \in \mathcal{M}_{i}$, the central server maintains a unit-size buffer, whose state at time $t$ is denoted by an indicator variable $b_{j}^{i}(t)$, where $b_{j}^{i}(t)=1$ implies that, at time $t$, user $j$ has a job stored at the central server, and $b_{j}^{i}(t)=0$ otherwise. We assume that if a buffer already contains a job and a new job request arrives from the same user, the new request is dropped. Moreover, if a user belongs to $k$ different sets, the central server maintains $k$ distinct unit-size buffers for that user, one for each user group. 

At time $t$, an indicator variable $c_{j}^{i}(t)$ takes the value $1$ if the $i$th network serves a job of user $j\in\mathcal{M}_{i}$, and $0$ otherwise. We assume that completing any job requires at least one time slot in every machine network. Furthermore, the service (job-completion) time in the $i$th network is modeled as a random variable that is independent and identically distributed (i.i.d.) across different job requests and independent across different machine networks. We denote this service time by $Y_{i}$, which has probability mass function $\bm{f}^{i}$, i.e., $P(Y_{i}=k) = \bm{f}^{i}(k)$. Thus, a job assigned to the $i$th network is completed in $k$ slots with probability $\bm{f}^{i}(k)$. Since the processing times are i.i.d. over different jobs, $Y_{i}$ does not depend on the job index. 

We further assume that a machine completes a job at the \emph{end} of a time slot, and we represent this event with an indicator random variable $d_{j}^{i}(t)$. Specifically, $d_{j}^{i}(t)=1$ indicates that, at the end of slot $t$, a job corresponding to user $j \in \mathcal{M}_{i}$ is completed, whereas $d_{j}^{i}(t)=0$ otherwise. The evolution of $b_{j}^{i}(t)$ is then given by
\begin{align}
     b_{j}^{i}(t) = \min\{b_{j}^{i}(t-1) - d_{j}^{i}(t-1) + a_{j}^{i}(t), 1\}.
\end{align}

At time $t$, we denote the \emph{age of job} corresponding to user $j \in \mathcal{M}_{i}$ by $\Delta_{j}^{i}(t)$, defined as the duration for which the current job has remained in the corresponding job buffer up to time $t$. Formally,
\begin{align}
    u_{j}^{i}(t) &= \sup\{t' \mid t' < t,\ d_{j}^{i}(t') = 1\}, \label{eq:2}\\
    \bar{u}_{j}^{i}(t) &= \inf\{t' \mid u_{j}^{i}(t) < t' \leq t,\ a_{j}^{i}(t') = 1\}, \label{eq:3}\\
    \Delta_{j}^{i}(t) &=
    \begin{cases}
        t - \bar{u}_{j}^{i}(t), & \bar{u}_{j}^{i}(t) \neq 0,\\
        0, & \bar{u}_{j}^{i}(t) = 0.
    \end{cases}
\end{align}
In (\ref{eq:2}) and (\ref{eq:3}), we assume that the supremum and infimum of an empty set are both equal to $0$.

Formally, at each time $t$, we have the following constraint,
\begin{align}
   & \sum_{j=1}^{M_{i}} c_{j}^{i}(t) \leq \bar{M}_{i}, \quad \forall i \in\{1,2,\cdots,N\}, \label{eq:5} \\ & \sum_{i=1}^{N} \sum_{j=1}^{M_{i}} c_{j}^{i}(t) \leq \bar{M}.\label{eq:6}
\end{align}
To make our problem non-trivial, we assume that
\begin{align}
    \sum_{i=1}^{N}{\bar{M}_{i}} > \bar{M}, \quad \bar{M} > \bar{M}_{i}, \quad \forall i\in\{1,2,\cdots,N\}.
\end{align}
In our system, different types of jobs may have different priorities. To model this, we associate a weight with the age of each job. For the job of user $j \in \mathcal{M}_{i}$, we assign a weight $w_{j}^{i}$, where a larger weight corresponds to a higher priority. We consider the following optimization problem:
\begin{align}
    &\inf_{\pi\in\Pi} \limsup_{T\rightarrow\infty}\frac{1}{T}\sum_{t=1}^{T}\sum_{i=1}^{N}\sum_{j=1}^{M_{i}}\mathbb{E}_{\pi}\big[w_{j}^{i}\Delta_{j}^{i}(t)\big] \nonumber\\
     &\ \text{s.t.} \quad (\ref{eq:5}),\ (\ref{eq:6}), \quad \forall t\in\{1,2,\cdots\}, \label{eq:8}
\end{align}
where $\Pi$ denotes the set of all causal scheduling policies. Note that, in~(\ref{eq:8}), we implicitly assume the initial condition $\Delta_{j}^{i}(0)=1$ for all $j\in\mathcal{M}_{i}$ and $i\in\{1,2,\cdots,N\}$, thus, we do not state it explicitly in (\ref{eq:8}). For notational convenience, we denote
\begin{align}
    \Delta^{{\avg},\pi}
    = \limsup_{T\rightarrow\infty}\frac{1}{T}
      \sum_{t=1}^{T}\sum_{i=1}^{N}\sum_{j=1}^{M_{i}}
      \mathbb{E}_{\pi}\big[w_{j}^{i}\Delta_{j}^{i}(t)\big].
\end{align}

Due to the constraints in (\ref{eq:5}), a \textit{network-level service outage} may occur; e.g., at time $t$, if network $i$ is already serving $\bar{M}_{i}$ jobs when a new job request from $\mathcal{M}_{i}$ arrives at the server. Similarly, due to the constraint in (\ref{eq:6}), a \textit{server-level service outage} may arise when, at time $t$, the central server is already serving $\bar{M}$ jobs and a new job request arrives from any of the $N$ user sets. To handle such situations, we allow \textit{service preemption} in the \emph{preemptive-resume} sense described in Section~\ref{sec:intro}. When a job for user $j$ is preempted, the central server stores its progress in the same job buffer introduced earlier. The preemption decision is made solely by the central server at the beginning of each time slot. We denote a scheduling policy $\pi$ as
\begin{align}
 \pi = \left(\left(\left(c_{j}^{i}(t)\right)_{j=1}^{M_{i}}\right)_{i=1}^{N}\right)_{t=1}^{\infty}.
\end{align}

Suppose a job from user $j$ in $\mathcal{M}_{i}$ has been served for the past $x>0$ slots. At the beginning of slot $t$, the server preempts this job and stores it in the buffer. Further, assume that, at the beginning of slot $t+x_{1}$, where $x_{1}>0$, the server resumes processing this job. Then, at the end of slot $t+x_{1}$, the job completes with probability,
\begin{align}
    \frac{\bm{f}^{i}(x+1)}{\sum_{k=x+1}^{\infty}\bm{f}^{i}(k)}. \nonumber
\end{align}
We denote by $\bar{\Delta}_{j}^{i}(t)$ the amount of time the job corresponding to user $j\in \mathcal{M}_{i}$ has been processed up to time $t$ (not including $t$). Formally,
\begin{align}
 \bar{\Delta}_{j}^{i}(t) = \sum_{\tau = \bar{u}_{j}^{i}(t)}^{t-1} c_{j}^{i}(\tau).
\end{align}
At time $t$, we denote the state corresponding to user $j\in\mathcal{M}_{i}$ by $s_{j}^{i}(t)= (\Delta_{j}^{i}(t),\bar{\Delta}_{j}^{i}(t))$. Likewise, the state of network $i$ at time $t$ is $s_{i}(t)=(s_{j}^{i}(t))_{j=1}^{M_{i}}$, and the state of the whole system is $s(t) = (s_{i}(t))_{i=1}^{N}$. Under a policy $\pi$, at each time $t$ the central server observes $s(t)$ and $\big(\big(a_{j}^{i}(t)\big)_{j=1}^{M_{i}}\big)_{i=1}^{N}$ to decide which jobs to serve.

\section{Max-Weight Policy}\label{sec:max-weight}
In this section, we derive a max-weight policy by minimizing the one-step conditional Lyapunov drift, which at time $t$ is defined as,
\begin{align}\label{eq:67}
    \mathbb{E}\!\left[\mathcal{L}(s(t+1)) - \mathcal{L}(s(t)) \,\middle|\, s(t)\right],
\end{align}
where we use the Lyapunov function
\begin{align}
    \mathcal{L}(s(t)) = \sum_{i=1}^{N} \sum_{j \in \mathcal{I}_{i}(t)} w_{j}^{i}\Delta_{j}^{i}(t),
\end{align}
where $\mathcal{I}_{i}(t)$ denotes the set of users in group $i$ that have a job at the central server at time $t$. Note that
\begin{align}
    &\mathbb{E}[\mathcal{L}(s(t+1))|s(t)] = \Bigg[\sum_{i=1}^{N}\sum_{j\in\mathcal{I}_{i}(t)}(1-c_{j}^{i}(t)) w_{j}^{i} (\Delta_{j}^{i}(t)+1) \nonumber\\&
    \qquad \quad + c_{j}^{i}(t)w_{j}^{i} \frac{\sum_{k=2}^{\infty}\bm{f}^{i}(\bar{\Delta}_{j}^{i}(t)+k)}{\sum_{k=1}^{\infty}\bm{f}^{i}(\bar{\Delta}_{j}^{i}(t)+k)}(\Delta_{j}^{i}(t)+1)\Bigg].
\end{align}
We define,
\begin{align}
    \alpha_{j}^{i}(t) 
    \triangleq
    \frac{\sum_{k=2}^{\infty} \bm{f}^{i}\!\big(\bar{\Delta}_{j}^{i}(t) + k\big)}
         {\sum_{k=1}^{\infty} \bm{f}^{i}\!\big(\bar{\Delta}_{j}^{i}(t) + k\big)}.
\end{align}
Then the one-step Lyapunov drift is,
\begin{align}\label{eq:45}
  \sum_{i=1}^{N} \sum_{j \in \mathcal{I}_{i}(t)} 
       w_{j}^{i} \Big( 1 + c_{j}^{i}(t)\big(\alpha_{j}^{i}(t) - 1\big)\big(\Delta_{j}^{i}(t)+1\big) \Big).
\end{align}
The first term in (\ref{eq:45}) is independent of the scheduling decisions $c_{j}^{i}(t)$, for all $i\in\{1,2,\cdots,N\}$ and $j\in\mathcal{I}_{i}(t)$. Thus, minimizing the one-step Lyapunov drift is equivalent to maximizing,
\begin{align}
    \sum_{i=1}^{N} \sum_{j \in \mathcal{I}_{i}(t)} 
    c_{j}^{i}(t) w_{j}^{i}\big(\Delta_{j}^{i}(t) + 1\big)\big(1 - \alpha_{j}^{i}(t)\big).
\end{align}
Hence, to implement the proposed \emph{max-weight policy}, we solve the following optimization problem at each time $t$:
\begin{align}
\max_{\left(\left(c_{j}^{i}(t)\right)_{j=1}^{M_{i}}\right)_{i=1}^{N}} & \displaystyle \sum_{i=1}^{N}\sum_{j\in{\mathcal{I}}_{i}(t)} c_{j}^{i}(t) w_{j}^{i}\big(\Delta_{j}^{i}(t) + 1\big)\big(1 - \alpha_{j}^{i}(t)\big) \nonumber\\
    \text{s.t.} \quad & (\ref{eq:5}),\ (\ref{eq:6}). \label{eq:36}
\end{align}

\section{NGM Policy}\label{sec:net-gain}
The problem in (\ref{eq:8}) constitutes an RMAB problem, where each user $j\in\mathcal{M}_{i}$ is treated as an arm, and the state of that arm at time $t$ is denoted by $s_{j}^{i}(t)$. In this section, we study the NGM policy to solve the optimization problem in (\ref{eq:8}).

\subsection{Lagrangian Relaxation and Dual Decomposition}
The constraints (\ref{eq:5}) and (\ref{eq:6}) in (\ref{eq:8}) are per slot constraints, i.e., the central server must satisfy them at every time slot. We relax these constraints to expected  time average constraints, leading to the following problem,
\begin{align}
    \inf_{\pi\in\Pi} & \limsup_{T\rightarrow\infty} \frac{1}{T} \sum_{t=1}^{T} \sum_{i=1}^{N}\sum_{j=1}^{M_{i}}\mathbb{E}_{\pi}[w_{j}^{i} \Delta_{j}^{i}(t)] \nonumber \\  
    \text{s.t.}& \limsup_{T\rightarrow\infty} \!\frac{1}{T} \sum_{t=1}^{T}\sum_{j=1}^{M_{i}}\! \mathbb{E}_{\pi}[c_{j}^{i}(t)] \leq \bar{M}_{i}, \ \forall{i}\in{1,2,\cdots,N} \nonumber \\ &\limsup_{T\rightarrow\infty} \frac{1}{T} \sum_{t=1}^{T} \sum_{i=1}^{N} \sum_{j=1}^{M_{i}} \mathbb{E}_{\pi}[c_{j}^{i}(t)] \leq \bar{M}.\label{eq:13}
\end{align}
We now employ $(N+1)$ Lagrange multipliers $\mu_{i}\geq0$ for $i\in\{1,2,\cdots,N\}$ and $\lambda\geq 0$ to form the following Lagrangian for the problem in (\ref{eq:13}),
\begin{align}
    &L(\pi,\bm{\mu},\lambda) = \limsup_{T\rightarrow\infty} \frac{1}{T}\sum_{t=1}^{T}\Bigg[\sum_{i=1}^{N}\sum_{j=1}^{M_{i} }\mathbb{E}_{\pi}[w_{j}^{i}\Delta_{j}^{i}(t) \nonumber\\&+\lambda c_{j}^{i}(t)] + \sum_{i=1}^{N}\mu_{i}\sum_{j=1}^{M_{i}} \mathbb{E}_{\pi}[c_{j}^{i}(t)]\Bigg] - \lambda \bar{M} - \sum_{i=1}^{N} \mu_{i}\bar{M}_{i},
\end{align}
where $\bm{\mu}= (\mu_{1},\mu_{2},\cdots,\mu_{N})$. Now, the dual problem of (\ref{eq:13}) is given by,
\begin{align}\label{eq:n16}
    \sup_{\lambda\geq 0,\bm{\mu}\geq \bm{0}} g(\lambda,\bm{\mu}),
\end{align}
where $g(\lambda,\bm{\mu})$ is the dual function, defined as,
\begin{align}\label{eq:16}
    g(\lambda,\bm{\mu}) = \inf_{\pi\in\Pi} L(\pi,\lambda,\bm{\mu}).
\end{align}
For the problem in (\ref{eq:16}), the term $-\lambda \bar{M} - \sum_{i=1}^{N}\mu_{i}\bar{M}_{i}$, is independent of $\pi$, thus, we omit it while solving (\ref{eq:16}).

Notice that the problem in (\ref{eq:16}) is linearly separable across networks and users. Thus, for a given $\bm{\mu}$ and $\lambda$, we decompose the problem in (\ref{eq:16}) into $\sum_{i=1}^{N}M_{i}$ sub-problems. The sub-problem corresponding to user  $j\in\mathcal{M}_{i}$ is,
\begin{align}\label{eq:18}
    \inf_{\pi_{j}^{i}\in\Pi_{j}^{i}} \limsup_{T\rightarrow\infty} \frac{1}{T}\sum_{t=1}^{T} \mathbb{E}_{\pi}[w_{j}^{i}\Delta_{j}^{i}(t) + (\lambda+\mu_{i})c_{j}^{i}(t)],
\end{align}
where $\pi_{j}^{i} = (c_{j}^{i}(t))_{t=1}^{\infty}$ and $\Pi_{j}^{i}$ is the set of all causal scheduling policies for user $j\in\mathcal{M}_{i}$. 

\subsection{Optimal Solution of (\ref{eq:18}) and Optimal Solution of (\ref{eq:n16})}\label{subsec:b}
In this subsection, we first find an optimal solution $\bar{\pi}_{j}^{i}$ for (\ref{eq:18}). 
Note that (\ref{eq:18}) is an average-cost Markov decision process (MDP), which we denote with $\zeta_{j}^{i}$. We first introduce the components of this MDP. A state of $\zeta_{j}^{i}$ is denoted by $s_{j}^{i}=(\Delta_{j}^{i},\bar{\Delta}_{j}^{i})$, where $\Delta_{j}^{i}$ denotes the age of job corresponding to user $j\in\mathcal{M}_{i}$, and $\bar{\Delta}_{j}^{i}$ denotes the total processing time for that job. If the job buffer for user $j\in\mathcal{M}_{i}$ is empty, we set $\bar{\Delta}_{j}^{i}=\infty$. We denote the state space of $\zeta_{j}^{i}$ with $\mathcal{S}_{j}^{i}$,
\begin{align}
    \mathcal{S}_{j}^{i} \!=\! \{(\Delta_{j}^{i},\bar{\Delta}_{j}^{i})| (\Delta_{j}^{i},\bar{\Delta}_{j}^{i}) \in \mathbb{N}^{2} \}\!\cup\!\{(0,\infty)\}.
\end{align}
We denote an action in $\zeta_{j}^{i}$ with $c_{j}^{i}\in\{0,1\}$, where $c_{j}^{i}=1$ indicates that the server serves user $j\in\mathcal{M}_{i}$, and $c_{j}^{i}=0$ otherwise. If $\bar{\Delta}_{j}^{i}$ is empty then only action $c_{j}^{i}=0$ is valid. For state $s_{j}^{i}$, we denote the set of valid actions with $\mathcal{A}_{j}^{i}(s_{j}^{i})$. The cost for state action pair $(s_{j}^{i},c_{j}^{i})$ is denoted by, 
\begin{align}
    {C}_{j}^{i}(s_{j}^{i},c_{j}^{i}) = w_{j}^{i}\Delta_{j}^{i} + (\lambda+\mu_{i}) c_{j}^{i}.
\end{align}
We denote the state transition probability from state $s_{j}^{i}$ to state $\bar{s}_{j}^{i}$ under action $c_{j}^{i}$ with $P_{j}^{i}(s_{j}^{i},\bar{s}_{j}^{i};c_{j}^{i})$. Next, we describe all possible non-zero state transition probabilities,
\begin{align}
    &P_{j}^{i}((0,\infty),(0,\infty);0) = 1-p_{j}^{i},\\&P_{j}^{i}((0,\infty),(0,0);0) = p_{j}^{i}, \\ & P_{j}^{i} ((\Delta_{j}^{i},\bar{\Delta}_{j}^{i}), (\Delta_{j}^{i}+1,\bar{\Delta}_{j}^{i});0) = 1, \\&P_{j}^{i}((\Delta_{j}^{i},\bar{\Delta}_{j}^{i}), (0,0);1) =  \frac{p_{j}^{i}\bm{f}^{i}(\bar{\Delta}_{j}^{i}+1)}{\sum_{k=1}^{\infty}\bm{f}^{i}(\bar{\Delta}_{j}^{i}+k)},\\ & P_{j}^{i}((\Delta_{j}^{i},\bar{\Delta}_{j}^{i}), (0,\infty);1) \!=\! \frac{(1\!-\!p_{j}^{i}) \bm{f}^{i}(\bar{\Delta}_{j}^{i}+1)}{\sum_{k=1}^{\infty}\bm{f}^{i}(\bar{\Delta}_{j}^{i}+k)},\\ &P_{j}^{i}((\Delta_{j}^{i},\bar{\Delta}_{j}^{i}), (\Delta_{j}^{i}\!+\!1,\bar{\Delta}_{j}^{i}\!+\!1);1) \!=\!\frac{ \sum_{k=2}^{\infty}\bm{f}^{i}(\bar{\Delta}_{j}^{i}\!+\!k)}{\sum_{k=1}^{\infty}\bm{f}^{i}(\bar{\Delta}_{j}^{i}\!+\!k)}.
\end{align}
Now, we employ relative value iteration to find an optimal solution of $\zeta_{j}^{i}$. Consider the following iteration on $n$,
\begin{align}\label{eq:24}
    &V_{j}^{i(n)}(s_{j}^{i};\lambda,\bm{\mu}) = \min_{c_{j}^{i}\in\mathcal{A}_{j}^{i}(s_{j}^{i})}\Big\{{C}_{j}^{i}(s_{j}^{i},c_{j}^{i})\!+\!\!\sum_{\bar{s}_{j}^{i}\in\mathcal{S}_{j}^{i}}P_{j}^{i}(s_{j}^{i},\bar{s}_{j}^{i};c_{j}^{i}) \nonumber\\
    &\qquad \qquad \ V_{j}^{i(n-1)}(\bar{s}_{j}^{i};\lambda,\bm{\mu})\Big\} - V_{j}^{i(n-1)} ((0,0);\lambda,\bm{\mu}),
\end{align}
where $V_{j}^{i(0)}(s_{j}^{i};\lambda,\bm{\mu})=0$, for all $s_{j}^{i}\in\mathcal{S}_{j}^{i}$. Similar to the literature \cite{banerjee2024preempt,banerjee2023re}, in computing (\ref{eq:24}), we bound $\Delta_{j}^{i}$ and $\bar{\Delta}_{j}^{i}$, by a natural number $Y_{\max}$, thereby bounding the state space $\mathcal{S}_{j}^{i}$. For clarity, we denote this bounded state space with $\mathcal{S}_{j}^{i}(Y_{\max})$. For this bounded state space, from \cite{bertsekas2012dynamic}, we know that (\ref{eq:24}) converges. We denote the relative value function obtained at convergence by $\bar{V}_{j}^{i}(s_{j}^{i};\lambda,\bm{\mu})$. We define
\begin{align}\label{eq:23}
    \bar{V}_{j}^{i}(s_{j}^{i};c_{j}^{i},\lambda,\bm{\mu}) = & {C}_{j}^{i}(s_{j}^{i},c_{j}^{i})\nonumber\\&
    +\sum_{\bar{s}_{j}^{i}\in\mathcal{S}_{j}^{i}}P_{j}^{i}(s_{j}^{i},\bar{s}_{j}^{i};c_{j}^{i}) \bar{V}_{j}^{i}(\bar{s}_{j}^{i};\lambda,\bm{\mu}).
\end{align}
Solving the minimization on the right-hand side of (\ref{eq:24}) for $\bar{V}_{j}^{i}(s_{j}^{i};\lambda,\bm{\mu})$, we obtain an optimal action for state $s_{j}^{i}\in\mathcal{S}_{j}^{i}(Y_{\max})$, which we denote by $\bar{c}_{j}^{i}(s_{j}^{i};\lambda,\bm{\mu})$. Next, we evaluate,
\begin{align}
    &\sum_{i=1}^{N}\sum_{j=1}^{M_{i}}\limsup_{T\rightarrow\infty}\left[\frac{1}{T}\sum_{t=1}^{T}\mathbb{E}[\bar{c}_{j}^{i}(s_{j}^{i}(t);\lambda,\bm{\mu})]\right] ,\label{eq:n24} \\ &\sum_{j=1}^{M_{i}}\left[ \limsup_{T\rightarrow\infty} \frac{1}{T} \sum_{t=1}^{T} \mathbb{E}[\bar{c}_{j}^{i}(s_{j}^{i}(t);\lambda,\bm{\mu})]\right], \ \forall{i}\in\{1,2,\cdots,N\}.\label{eq:25}
\end{align}
To this end, we employ the following relative value iteration:
\begin{align}\label{eq:26}
    \tilde{V}_{j}^{i(n)}(s_{j}^{i};&\lambda,\bm{\mu})= \tilde{C}(s_{j}^{i},\bar{c}_{j}^{i}(s_{j}^{i};\lambda,\bm{\mu})) \nonumber\\
    &+\sum_{\bar{s}_{j}^{i}\in\mathcal{S}_{j}^{i}}P_{j}^{i} (s_{j}^{i},\bar{s}_{j}^{i};\bar{c}_{j}^{i}(s_{j}^{i};\lambda,\bm{\mu})) 
     \tilde{V}_{j}^{i(n-1)} (\bar{s}_{j}^{i};\lambda,\bm{\mu}) \nonumber\\& - \tilde{V}_{j}^{i(n-1)} ((0,0);\lambda,\bm{\mu}),
\end{align}
with initialization $\tilde{V}_{j}^{i(0)}(s_{j}^{i})=0$, for all $s_{j}^{i}\in\mathcal{S}_{j}^{i}(Y_{\max})$. In (\ref{eq:26}), we used,
\begin{align}
    \tilde{C}(s_{j}^{i},\bar{c}_{j}^{i}(s_{j}^{i},\lambda,\bm{\mu})) = \begin{cases}
        1, &\bar{c}_{j}^{i}(s_{j}^{i},\lambda,\bm{\mu})=1, \\ 0, &\bar{c}_{j}^{i}(s_{j}^{i},\lambda,\bm{\mu})=0.
    \end{cases}
\end{align}
We denote the relative value function obtained at convergence of (\ref{eq:26}) by $\bar{\tilde{V}}_{j}^{i}(s_{j}^{i};\lambda,\bm{\mu})$. According to \cite{bertsekas2012dynamic} we have, 
\begin{align}
    \bar{\tilde{V}}_{j}^{i}((0,0);\lambda,\bm{\mu}) = \lim_{T\rightarrow\infty}\frac{1}{T}\sum_{t=1}^{T}\mathbb{E}[\bar{c}_{j}^{i}(s_{j}^{i}(t);\lambda,\bm{\mu})]. 
\end{align}
Thus, we have (\ref{eq:n24}) and (\ref{eq:25}). 

To obtain the optimal Lagrangian multipliers $(\lambda^{*},\bm{\mu}^{*})$ in (\ref{eq:n16}), we employ a sub-gradient ascent algorithm. The $(k+1)$th update of $(\lambda,\bm{\mu})$, denoted $(\lambda^{(k+1)},\bm{\mu}^{(k+1)})$, is given by
\begin{align}
    &\lambda^{(k+1)} 
    \!\!=\!\!\left[\!\lambda^{(k)}\! \!+ \!\tau(k)\!\left(\sum_{i=1}^{N}\sum_{j=1}^{M_{i}}\bar{\tilde{V}}_{j}^{i}((0,0);\lambda^{(k)},\bm{\mu}^{(k)}) \!- \!\bar{M}\!\right) \!\right]_{\!+}\!\!, \\&
    \mu_{i}^{(k+1)} \!=\! \left[\mu_{i}^{(k)}\! +\! \bar{\tau}_{i}(k)\left(\sum_{j=1}^{M_{i}}\bar{\tilde{V}}_{j}^{i}((0,0);\lambda^{(k)},\bm{\mu}^{(k)}) \!-\!\bar{M}_{i}\right)\right]_{\!+} \nonumber\\& \qquad \qquad\qquad\qquad\qquad\qquad\qquad\forall{i}\in\{1,2,\cdots,N\},
\end{align}
where $[x]_{+}=\max\{x,0\}$. We choose the step sizes as,
\begin{align}
    \tau(k) = \frac{h}{\sqrt{k}}, \qquad \bar{\tau}_{i}(k) = \frac{\bar{h}_{i}} {\sqrt{k}},
\end{align}
where $h$ and $(\bar{h}_{i})_{i=1}^{N}$ are positive parameters.

\subsection{Net-Gain Maximization Preemptive Scheduling Policy}
At time $t$, if the job buffer corresponding to user $j\in\mathcal{M}_{i}$ is empty, the server does not schedule that user. Hence, we only determine scheduling decisions for state $s_{j}^{i}\in\mathcal{S}_{j}^{i}$ with $\bar{\Delta}_{j}^{i}\neq \infty$. For the optimal Lagrangian multipliers $(\lambda^{*},\bm{\mu}^{*})$, obtained in Subsection~\ref{subsec:b}, and for all $s_{j}^{i}\in\mathcal{S}_{j}^{i}(Y_{\max})$ such that $\bar{\Delta}_{j}^{i}\neq \infty$, we compute, 
\begin{align}\label{eq:32}
    \beta_{j}^{i}(s_{j}^{i})=\bar{V}_{j}^{i}(s_{j}^{i};0,\lambda^{*},\bm{\mu}^{*}) - \bar{V}_{j}^{i} (s_{j}^{i};1,\lambda^{*},\bm{\mu}^{*}).
\end{align}
For state $s_{j}^{i}$, the quantity $\beta_{j}^{i}(s_{j}^{i})$ may be interpreted as the \textit{net-gain} of employing the \textit{active action} ($c_{j}^{i}=1$) relative to the \textit{passive action} ($c_{j}^{i}=0$). Now, consider that at time $t$, for user $j\in\mathcal{M}_{i}$ and $j'\in\mathcal{M}_{i'}$ we have $\beta_{j}^{i}(s_{j}^{i}) > \beta_{j'}^{i'}(s_{j'}^{i'})$, then, intuitively, the central server should prioritize scheduling user $j\in$ $\mathcal{M}_{i}$, i.e., employing active action, over user $j'\in\mathcal{M}_{i'}$. Similarly, for two users $j,j'\in\mathcal{M}_{i}$, if $\beta_{j}^{i}(s_{j}^{i})>\beta_{j'}^{i}(s_{j'}^{i})$, then, intuitively, the central server should again prioritize user $j$ over $j'$. The NGM preemptive scheduling policy formalizes this intuition and systematically selects users based on their net-gain values. To implement the NGM policy, at every slot $t$, the server solves the same optimization problem as in~(\ref{eq:36}), replacing the objective function in (\ref{eq:36}) with $   \sum_{i=1}^{N}\sum_{j\in{\mathcal{I}}_{i}(t)} c_{j}^{i}(t)\beta_{j}^{i}(s_{j}^{i}(t))$. 

Note that the optimal Lagrangian multipliers, $(\lambda^{*},\bm{\mu}^{*})$, obtained in Subsection~\ref{subsec:b}, are computed with respect to the state space $\mathcal{S}_{j}^{i}(Y_{\max})$. We choose $Y_{\max}$ sufficiently large such that at time $t$, the probability that $s_{j}^{i}(t)\notin\mathcal{S}_{j}^{i}(Y_{\max})$ is small. However, we cannot guarantee that $s_{j}^{i}(t)$ always remains in $\mathcal{S}_{j}^{i}(Y_{\max})$. If at time $t$, $s_{j}^{i}(t)\notin\mathcal{S}_{j}^{i}(Y_{\max})$, we select a large enough $Y_{\max}'$, such that $s_{j}^{i}(t)\in\mathcal{S}_{j}^{i}(Y_{\max}')$. Using the same multipliers $(\lambda^{*},\mu^{*})$ previously computed for $\mathcal{S}_{j}^{i}(Y_{\max})$, we run the relative value iteration in (\ref{eq:24}) on $\mathcal{S}_{j}^{i}(Y_{\max}')$, and evaluate the corresponding relative value function. With this updated value function, we compute the net-gain for state $s_{j}^{i}(t)$ using (\ref{eq:32}), and solve the optimization problem in (\ref{eq:36}), replacing the objective function with $   \sum_{i=1}^{N}\sum_{j\in{\mathcal{I}}_{i}(t)} c_{j}^{i}(t)\beta_{j}^{i}(s_{j}^{i}(t))$. 

\section{Whittle Index Policy for Geometric Job Completion}\label{sec:whittle}
In this section, we study the Whittle index policy for our problem under the assumption that the job completion time of the $i$th network $Y_{i}$ follows a geometric distribution with success probability $q_{i}$, for all $i\in\{1,2,\cdots,N\}$. We begin by focusing on the optimization problem for user $j\in\mathcal{M}_{i}$ given in (\ref{eq:18}). The MDP formulation in Subsection~\ref{subsec:b} remains valid for geometric service time. However, due to the memoryless property of the geometric distribution, when the server has a job for user $j\in\mathcal{M}_{i}$, an optimal action for state $s_{j}^{i}$ in the MDP of (\ref{eq:18}) becomes independent of the state variable $\bar{\Delta}_{j}^{i}$. Consequently, we use $\bar{\Delta}_{j}^{i}$ only as an indicator to denote whether a job is present at the server for user $j\in\mathcal{M}_{i}$. We set $\bar{\Delta}_{j}^{i}=0$ if a job exists, and $\bar{\Delta}_{j}^{i}=\infty$ otherwise.

Similar to Section~\ref{sec:net-gain}, we refer to $c_{j}^{i}=1$ as the active action and $c_{j}^{i}=0$ as the passive action. From (\ref{eq:18}), we see that taking active action for user $j\in\mathcal{M}_{i}$ incurs a scheduling cost or activation cost $\lambda+\mu_{i}$, while the passive action does not incur any such cost. For notational convenience, we denote $\lambda+\mu_{i}$ with $\bar{\lambda}_{i}$. For a given $\bar{\lambda}_{i}$, we define the set of all passive states, i.e., states for which passive action is optimal, as $\mathcal{P}_{j}^{i}(\bar{\lambda}_{i})$. Formally, we define it as,
\begin{align}
    \mathcal{P}_{j}^{i}(\bar{\lambda}_{i})
    =&\Big\{s_{j}^{i}\in\mathcal{S}_{j}^{i}\,:\,\bar{\Delta}_{j}^{i}=0,\ 
    \bar{V}_{j}^{i}(s_{j}^{i};0,\bar{\lambda}_{i}) \le \bar{V}_{j}^{i}(s_{j}^{i};1,\bar{\lambda}_{i})
    \Big\} \nonumber\\
    &\cup \Big\{s_{j}^{i}\in\mathcal{S}_{j}^{i}\,:\,\bar{\Delta}_{j}^{i}=\infty\Big\},
\end{align}
where $\bar{V}_{j}^{i}(s_{j}^{i};c_{j}^{i},\bar{\lambda}_{i})$ is the same as $\bar{V}_{j}^{i}(s_{j}^{i};c_{j}^{i},\lambda,\bm{\mu})$, defined in (\ref{eq:23}), evaluated with $\bar{\lambda}_{i}=\lambda+\mu_{i}$. We say that user $j\in\mathcal{M}_{i}$ is indexable if the set of passive states expands monotonically with $\bar{\lambda}_{i}$, i.e., for $\bar{\lambda}_{i,1}\leq \bar{\lambda}_{i,2}$, we have $\mathcal{P}_{j}^{i}(\bar{\lambda}_{i,1})\subseteq \mathcal{P}_{j}^{i}(\bar{\lambda}_{i,2})$ and $\mathcal{P}_{j}^{i}(\infty)=\mathcal{S}_{j}^{i}$. If for every network $i\in\{1,2,\cdots,N\}$, all users $j\in\mathcal{M}_{i}$ are indexable, then we say that the scheduling problem in (\ref{eq:8}) is indexable. 

In Theorem~\ref{thm1}, we establish the indexability of the scheduling problem defined in (\ref{eq:8}). 
\begin{theorem}\label{thm1}
    The scheduling problem in (\ref{eq:8}) is indexable.
\end{theorem}

Now, we obtain the Whittle index for a state $(\Delta_{j}^{i},0)$. First, we give a formal definition of the Whittle index.
\begin{definition}
    For a state $s_{j}^{i}=(\Delta_{j}^{i},0)$, the Whittle index is,
    \begin{align}
        W_{j}^{i}(s_{j}^{i})= \inf\{\bar{\lambda}_{i}|s_{j}^{i}\in\mathcal{P}_{j}^{i}(\bar{\lambda}_{i})\}.
    \end{align}
\end{definition}

In Theorem~\ref{thm2}, we provide the Whittle index for states~in~$\mathcal{S}_{j}^{i}$.
\begin{theorem}\label{thm2}
    For a state $s_{j}^{i}=(\Delta_{j}^{i},0)$, the Whittle index is,
    \begin{align}
        W_{j}^{i}(s_{j}^{i})= w_{j}^{i} \left(\frac{q_{i} (\Delta_{j}^{i})^{2}}{2} + \left(1\!-\!\frac{q_{i}}{2} \!+\!\frac{q_{i}}{p_{j}^{i}}\right)\Delta_{j}^{i}+\frac{1}{p_{j}^{i}} \right).
    \end{align}
\end{theorem}

Now, analogous to Section~\ref{sec:max-weight}, to implement the Whittle index policy, at every slot $t$, the server solves the optimization problem in (\ref{eq:36}), replacing the objective function with $   \sum_{i=1}^{N}\sum_{j\in{\mathcal{I}}_{i}(t)} c_{j}^{i}(t)W_{j}^{i}(s_{j}^{i}(t))$.

\section{WIMWF Policy}\label{sec:max-aid-whittle}
In Section~\ref{sec:whittle}, we derived the Whittle index policy for geometric service times. For general service times, however, we do not have the indexability property. Motivated by the Whittle index policy, we therefore propose a \emph{Whittle index-like policy} that does not require indexability. For a fixed state $s_{j}^{i} = (\Delta_{j}^{i}, \bar{\Delta}_{j}^{i})\in\mathcal{S}_{j}^{i}(Y_{\max})$, with $\bar{\Delta}_{j}^{i} \neq \infty$, we define,
\begin{align}
    \phi(\bar{\lambda}_{i})
    \triangleq
    \bar{V}_{j}^{i}(s_{j}^{i};0,\bar{\lambda}_{i})
    -
    \bar{V}_{j}^{i}(s_{j}^{i};1,\bar{\lambda}_{i}),
\end{align}
where similar to Section~\ref{sec:whittle}, we write $\bar{\lambda}_{i} = \lambda + \mu_{i}$ for simplicity. In Lemma~\ref{lemma:1}, we show that $\phi(\bar{\lambda}_{i})$ is a continuous function of $\bar{\lambda}_{i}$.

\begin{lemma}\label{lemma:1}
    The function $\phi(\bar{\lambda}_{i})$ is a concave and continuous function of $\bar{\lambda}_{i}\in (0,\infty)$. 
\end{lemma}

With technique similar to that of \cite{banerjee2024preempt, banerjee2024wiopt}, we can prove Lemma~\ref{lemma:1}. We aim to find a value $\bar{\lambda}_{i}^{\mathrm{op}}$, such that, we have $\phi(\bar{\lambda}_{i}^{\mathrm{op}})=0$. For each candidate $\bar{\lambda}_{i}$, we evaluate $\phi(\bar{\lambda}_{i})$ via relative value iteration \cite{bertsekas2012dynamic}. If such a root, $\bar{\lambda}_{i}^{\mathrm{op}}$, exists, we find it by bisection search and denote it by $\bar{W}_{j}^{i}(s_{j}^{i})$. Note that, for a state $s_{j}^{i}$, if such a root does not exist or we fail to obtain it, we instead define,
\begin{align}\label{eq:73}
    \bar{W}_{j}^{i}(s_{j}^{i})
    \triangleq
    \frac{w_{j}^{i} (\Delta_{j}^{i}+1)\,\bm{f}^{i}(\bar{\Delta}_{j}^{i}+1)}
         {\sum_{k=1}^{\infty} \bm{f}^{i}(\bar{\Delta}_{j}^{i}+k)},
\end{align}
i.e., we use the weight derived in Section~\ref{sec:max-weight} by minimizing the one-slot Lyapunov drift.

Since we do not establish indexability in the case of general service times, the outcome of the bisection search may depend on the choice of initial search interval and may yield different values of $\bar{W}_{j}^{i}(s_{j}^{i})$. For concreteness, we now describe the initial search interval, $[\bar{\lambda}_{i}^{L},\bar{\lambda}_{i}^{U}]$, used in this work. 

We begin by setting $\bar{\lambda}_{i}^{L}=0$ and evaluating $\phi(0)$. If $\phi(0)\leq 0$, we simply set $\bar{W}_{j}^{i}(s_{j}^{i})=0$. Otherwise, when $\phi(0)>0$, we choose a heuristic guess for $\bar{\lambda}_{i}^{U}$,
\begin{align}
    \bar{\lambda}_{i}^{U}=\max\{1,\, w_{j}^{i}(\Delta_{j}^{i}-1)\},
\end{align}
and evaluate $\phi(\bar{\lambda}_{i}^{U})$. If $\phi(\bar{\lambda}_{i}^{U})>0$, we repeatedly double $\bar{\lambda}_{i}^{U}$ (i.e., $\bar{\lambda}_{i}^{U}\leftarrow 2\bar{\lambda}_{i}^{U}$) and recompute $\phi(\bar{\lambda}_{i}^{U})$ until we find a value $\bar{\lambda}_{i}^{U}$ with $\phi(\bar{\lambda}_{i}^{U})\leq 0$. The last value with $\phi(\bar{\lambda}_{i})>0$ is taken as $\bar{\lambda}_{i}^{L}$, which yields an interval satisfying $\phi(\bar{\lambda}_{i}^{L})>0$ and $\phi(\bar{\lambda}_{i}^{U})\leq 0$. From Lemma~\ref{lemma:1}, we say that the bisection search converges on the interval $[\bar{\lambda}_{i}^{L},\bar{\lambda}_{i}^{U}]$, and provides us $\bar{\lambda}_{i}^{\mathrm{op}}=\bar{W}_{j}^{i}(s_{j}^{i})$. Note that, if we are unable to find a $\bar{\lambda}_{i}^{ U}$ such that $\phi(\bar{\lambda}_{i}^{U})\leq 0$, then we cannot employ the bisection search; thus, in this case, we use (\ref{eq:73}) as the index. If at time $t$, we have $s_{j}^{i}(t) \notin \mathcal{S}_{j}^{i}(Y_{\max})$, we again use (\ref{eq:73}) to compute the index.

Finally, to implement this policy, at each time $t$, we solve the same optimization problem as in~(\ref{eq:36}), replacing the objective function in (\ref{eq:36}) with $\sum_{i=1}^{N} \sum_{j\in{\mathcal{I}}_{i}(t)} c_{j}^{i}(t)
\bar{W}_{j}^{i}(s_{j}^{i}(t))$. 

\section{Numerical Results}\label{sec:num_res}
In this section, we will examine how the proposed policies behave as the number of users grows asymptotically. We use the term asymptotically in the following sense. We start from a {base system} with $N=3$ networks, each serving $M_{i}=3$ users. The central server can schedule at most $\bar{M}=2$ jobs per slot overall, and each network $i$ can serve at most $\bar{M}_{i}=2$ jobs per slot. The job arrival probabilities are, 
\begin{align}
   \bm {p} =
    \begin{bmatrix}
        0.3 & 0.4 & 0.5 \\
        0.4 & 0.2 & 0.6 \\
        0.1 & 0.3 & 0.2
    \end{bmatrix},
\end{align}
where the $(i,j)$ entry corresponds to $p_{j}^{i}$. The corresponding weights are,
\begin{align}
    \bm{w}
    =
    \begin{bmatrix}
        2 & 3 & 1 \\
        1 & 2 & 3 \\
        4 & 1 & 2
    \end{bmatrix},
\end{align}
where the $(i,j)$ entry corresponds to $w_{j}^{i}$.

For an integer scaling parameter $r \geq 1$, we construct a larger system by creating $r$ independent copies of this base system. For each original network $i$ and its three users, we introduce $r$ identical replicas with the same arrival probabilities, weights, service time distribution $\bm{f}^{i}$, and per-network capacity $\bar{M}_{i}$. The global scheduling capacity of the central server scales linearly with $r$, i.e., it becomes $r\bar{M}$, while the per-network capacity of each individual copy remains $\bar{M}_{i}$. Under this construction, both the number of networks and the total number of users grow linearly with $r$, while the local structure of each network and user group remains unchanged. This notion of an asymptotic regime is similar in spirit to that used in papers establishing the asymptotic optimality of the Whittle index; see, for example,~\cite{verlupewhittle}.

By varying $r$ and plotting the \emph{ormalized long-term average weighted age of job}, we observe how each policy $\pi$ behaves as the number of users (and networks) becomes large. Specifically, we obtain the normalized long-term average weighted age of job by dividing $\Delta^{\mathrm{avg},\pi}$ with the scaling parameter $r$.

First, we consider a non-geometric service time distribution,
\begin{align}
    \bm{f} =
    \begin{bmatrix}
        0.1 & 0.2 & 0.1 & 0.1 & 0.4 & 0.1 \\
        0.2 & 0.3 & 0.5 & 0   & 0   & 0   \\
        0.1 & 0.5 & 0.3 & 0.1 & 0   & 0
    \end{bmatrix},
\end{align}
where the $i$th row corresponds to $\bm{f}^{i}$. We compare four policies:
(i) the Lyapunov-based max-weight (MWL) policy from Section~\ref{sec:max-weight};
(ii) a heuristic max-weight (MWH) policy  that uses the simpler weight $w_{j}^{i}\Delta_{j}^{i}$;
(iii) the NGM policy from  Section~\ref{sec:net-gain} ; and
(iv) the WIMWF policy from Section~\ref{sec:max-aid-whittle}. For the NGM policy, we set $Y_{\max}=10000$, while for the WIMWF policy, we use $Y_{\max}=50$. 

\begin{figure}[t]
    \centerline{\includegraphics[width = 0.9\columnwidth, height=0.72\columnwidth]{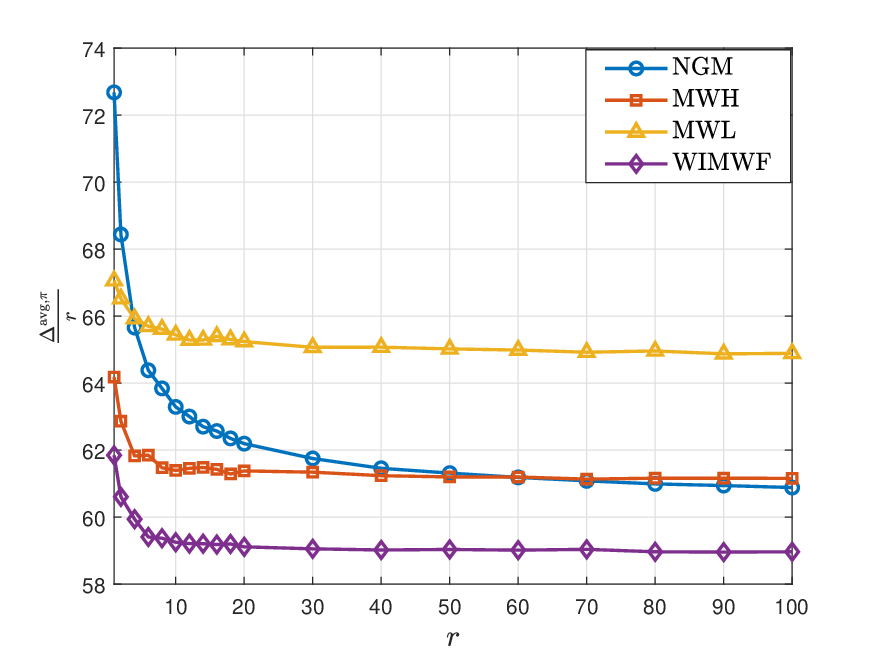}}
    \caption{Normalized average weighted age of job versus scaling factor $r$ for a general (non-geometric) service time distribution.}
    \label{fig:2}
\end{figure}

In Fig.~\ref{fig:2}, we see that for small values of $r$, the MWL and MWH policies achieve a lower $\frac{\Delta^{\mathrm{avg},\pi}}{r}$ than the NGM policy, reflecting their good performance in relatively small systems. However, as $r$ increases, the performance of the NGM policy improves steadily and eventually surpasses both MWL and MWH. Thus, we may regard the NGM policy as asymptotically better than the MWL and MWH policies. Across all values of $r$, WIMWF policy achieves the lowest normalized age among the four policies. Finally, we note that, in our simulation study we did not need to use the fallback weight in (\ref{eq:73}) to compute $\bar{W}_{j}^{i}(s_{j}^{i})$: the iterative procedure always converged, and for all $t$ we had $s_{j}^{i}(t)\in\mathcal{S}_{j}^{i}(Y_{\max})$. From Fig.~\ref{fig:2}, we also observe that MWH outperforms MWL. This indicates that minimizing the one-slot Lyapunov drift provides a systematic way to construct a max-weight policy; however, depending on the system, a heuristic max-weight policy can still outperform the one-slot drift–minimizing policy.

\begin{figure}[t]
    \centerline{\includegraphics[width = 0.9\columnwidth, height=0.72\columnwidth]{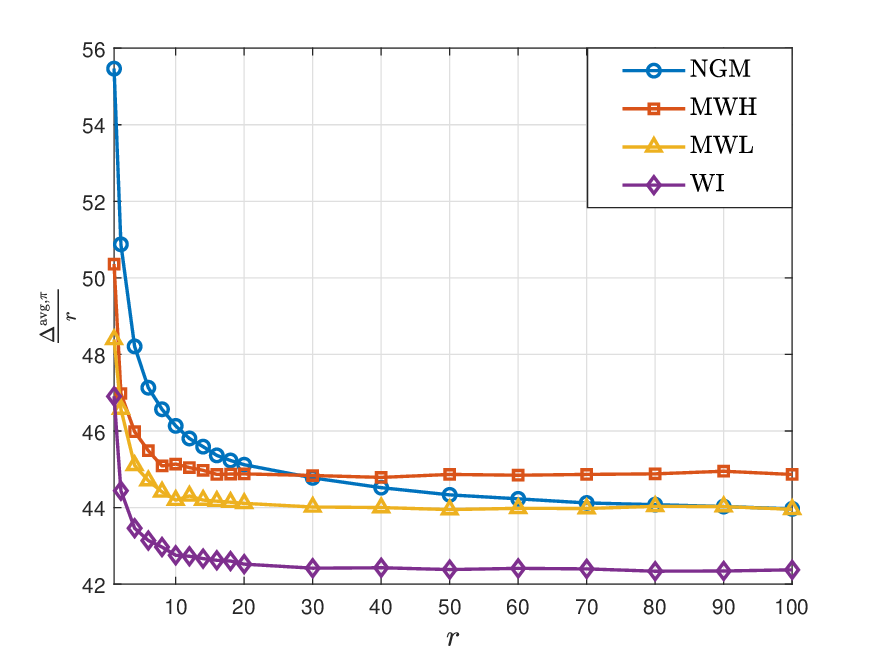}}
    \caption{Normalized average weighted age of job versus scaling factor $r$ for geometric service times.}
    \label{fig:3}
\end{figure}

Now, we focus on geometric service times, where the job completion probabilities of the three base networks are $0.3$, $0.5$, and $0.7$ for networks $1$, $2$, and $3$, respectively. Again, we compare four policies: the NGM policy, the MWH policy, the MWL policy, and the Whittle index (WI) policy from Section~\ref{sec:whittle}. For the NGM policy, we again set $Y_{\max} = 10000$. In Fig.~\ref{fig:3}, we see that for small values of $r$, the MWH and MWL policies outperform the NGM policy. As $r$ increases, however, the NGM policy catches up with the MWH and MWL policies and eventually surpasses them, again indicating that NGM is asymptotically better than max-weight policies. The results also show that the Whittle index policy consistently yields the lowest normalized age across all scaling factors $r$.

\section{Conclusion}
We studied a preemptive-resume job-assignment system in which a central server routes heterogeneous user jobs to task-specific networks of machines under both a global server-capacity constraint and per-network computation constraints. To capture the timeliness of the job completions, we introduced the \emph{age of job} metric and formulated the problem of minimizing the long-term weighted average age of job. We then proposed and analyzed several scheduling policies. First, we derived a Lyapunov-drift-based max-weight policy. Next, we formulated the problem as an RMAB and studied the NGM policy. For geometric job completion times, we established indexability and derived a closed-form Whittle index, leading to a Whittle index scheduling policy. For general service time distributions, where indexability is not guaranteed, we introduced a Whittle index-like policy WIMWF, that computes an equality point between active and passive actions, with a Lyapunov-based fallback weight when needed. Finally, we evaluated and compared all the proposed policies numerically.

\bibliographystyle{unsrt}
\bibliography{references}

\end{document}